# The Earth's oscillating electric field (T = 1 day) in relation to the occurrence time of large EQs (Ms ≥ 5.0R). A postulated theoretical physical working model and its statistical validation.

Thanassoulas[1], C., Klentos[2], V., Verveniotis, G.[3], Zymaris, N.[4]

1. Retired from the Institute for Geology and Mineral Exploration (IGME), Geophysical Department, Athens, Greece.
   e-mail: thandin@otenet.gr - URL: www.earthquakeprediction.gr

2. Athens Water Supply & Sewerage Company (EYDAP),
   e-mail: klenvas@mycosmos.gr - URL: www.earthquakeprediction.gr

3. Ass. Director, Physics Teacher at 2nd Senior High School of Pyrgos, Greece.
   e-mail: gver36@otenet.gr - URL: www.earthquakeprediction.gr

4. Retired, Electronic Engineer.

**Abstract**

The mechanically oscillating, due to tidal forces, lithosperic plate activates, because of its high content in quartzite, the generation of a piezoelectric field. Due to the same mechanical oscillation the lithosphere is generally at a state of an oscillating stress load. Therefore, large EQs which occur at the peaks of the stress load must coincide with the peaks of the generated piezoelectric potential. In this work a physical mechanism is postulated that accounts for the latter hypothesis. The postulated model is statistically tested by comparing the time of occurrence of **280** large EQs (Ms ≥ **5.0R**) which occurred during the period from **2003** to **2011**, to the same period of time Earth's electric field registered at **ATH** (Athens) and **PYR** (Pyrgos) monitoring sites located in Greece. The comparison has been made for the oscillating component of **T = 1 day** and for both the **E – W** and **N – S** directions. The statistical results indicate that the postulated model does not behave randomly. Instead, it represents a smooth normal distribution which peaks on the zero deviation in time between the time of occurrence of the large EQs and the amplitude peaks of the Earth's oscillating electric field. Therefore, the proposed physical model is an acceptable one and can be used for the finer refinement of the prediction of the occurrence time of a large EQ within a day's time period.

## 1. Introduction.

In the field of the earthquake prediction important advances has been achieved, in the past few decades of years, by the study of the Earth's electric field. Electric preseismic signals have been observed that are generated some time before large EQs. The observed preseismic electric signals have been attributed to various physical generating mechanisms while the same signals are represented by different type of "electric signature". Sobolev (1975) presented a bay-type very long period (of some months) preseismic signal, Varotsos et al. (1981) observed the generation of "train-like" electric short pulses preseismic electric signals while Thanassoulas et al. (1986, 1993) observed the generation of an oscillating electric precursor for a few days before the occurrence time of a large EQ. Detailed analyses and references for these preseismic electric precursors have been presented by Varotsos (2005) and Thanassoulas (2007). A common feature of all previously referred "short" preseismic electric signals is that all of them are superposed on the Earth's regional – background electric field, and can be directly observed and be registered by any appropriate electric potential registering apparatus.

In the recent last few years a different type of processing of the Earth's electric field revealed that it can provide us with valuable information that concerns the prediction of large EQs. Seismic electric precursors are indirectly revealed by appropriate processing of the Earth's regional electric field. The large scale lithosperic piezoelectricity model suggested by Thanassoulas et al. (1986, 1993) was again verified by recent studies (Thanassoulas, 2008, 2008b). It was demonstrated that the regional Earth's electric field, long before the occurrence time of a large EQ, in most cases, takes the form of a typical "strain – electric potential" curve which is characteristic for the piezoelectric materials while the large EQ (rock fracturing) takes place during the upper non-linear part of that curve following the rock mechanics laws (Jaeger, 1974). A quite different seismic electric precursor is the "strange attractor like" one (Thanassoulas 2007; Thanassoulas et al. 2008a; 2009; 2009a). In this type of analysis, the registered, by two distant monitoring sites, Earth's electric fields and specifically its oscillating components are combined to generate phase maps. It has been observed that ellipses are generated short before the occurrence time of a large EQ in contrast to the presence of hyperbolas when there is no adequate strain load in order to trigger in the near future a large EQ.

All the previously referred seismic electric precursors have been observed (or revealed by processing) in a large data set of electric field registrations by three monitoring sites located in Athens (**ATH**), Pyrgos (**PYR**) and Hios Island (**HIO**) in Greece. The specific data set spans for more than eight (**8**) years for the **ATH** and **PYR** monitoring sites and for almost four (**4**) years for the **HIO** case. Scientists who are interested in the topic may download the entire data set from the link: www.earthquakeprediction.gr

During that long experiment and specifically during the updating of the referred internet site with the daily registered data, our attention was attracted from the fact that some times large EQs did coincide in occurrence time with the same day amplitude peaks of the Earth's oscillating (**T = 1 day**) electric field component. Therefore, it is interesting to study this coincidence along the longest (more than eight years) data sets obtained from **PYR** and **ATH** monitoring sites. A sample of ten (**10**) days recording of the Earth's raw electric field (**ATH**) and its corresponding oscillating component is presented in the following figure (**1**). The oscillating component has been obtained after band-pass (FFT) filtering applied on the raw data.

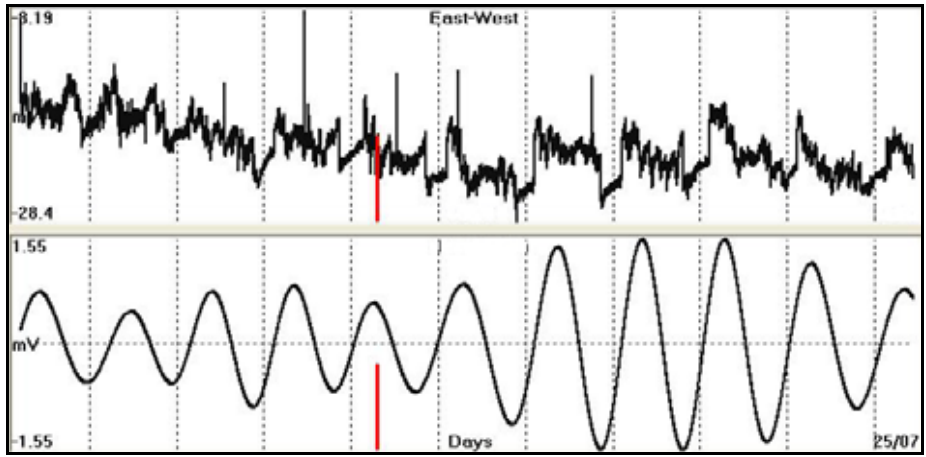

**Fig. 1.** Earth's electric field raw data (upper graph) and the corresponding oscillating (lower graph, **T = 1 day**) component. The red bar indicates the occurrence time of a large (**Ms ≥ 5.0R**) EQ. **ATH** monitoring site.

What actually is observed in **ATH** and **PYR** monitoring sites is (as a sample) presented in the following figure (**2**). The oscillating field of each monitoring site, for both **N – S** and **E – W** components, is compared to the occurrence time of the large (**Ms ≥ 5.0R**) EQs of the same period of time.

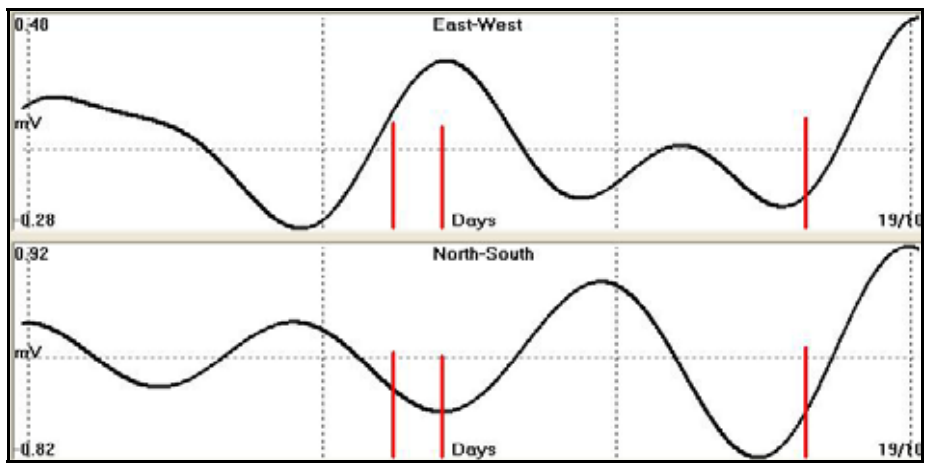

**Fig. 2.** Three days recording compared to the occurrence time of the same period large (**Ms ≥ 5.0R**) EQs. Left EQ: **dt = -235** minutes from **N – S** peak, Middle EQ: **dt = -1** minute from **N – S** peak, Right EQ: **dt = 116** minutes from **E – W** peak. **ATH** monitoring site.

Some of the large EQs occurred at a rather long time distance from the closest amplitude peak of the oscillating electric field while other occurred very close to it. In figure (**2**) the shortest observed time distance is **dt = -1** minute for the middle EQ while the longest one is **dt = -235** minutes. The time of occurrence of the large EQs can be either related or not, by any physical mechanism, to the Earth's oscillating electric field amplitude peaks. Which case is correct can be resolved by a simple statistical analysis applied on all the determined dt for all large EQs of the study period.

**2. Theoretical analysis - postulated model.**

A brief theoretical analysis is presented as follows, and as a flow-chart too, in figure (**3**):
The lithosphere (**block 1**) is set to a continuous mechanical oscillating mode due to the tidal forces which are applied upon it. Therefore, the stress load of the lithospheric plate oscillates accordingly. Thus, the stress load of the lithosphere can be expressed as a function (**f**) of the various tidal oscillating components:

$$\text{Lithospheric Stress Load} = f(T_1, T_2 \ldots\ldots T_n) \qquad (1)$$

Where **T1, T2, ……. Tn** are the periods of the tidal oscillating components.

Taking into account the fact that the lithosphere contains a large amount of quartzite it is logical to accept the activation of the piezoelectric mechanism (**block 2**). The generated piezoelectric potential is expressed as:

$$\text{Piezoelectric Potential} = K*(\text{Stress load}) \qquad (2)$$

Where (**K**) is the piezoelectric constant that characterizes the stress loaded lithospheric material.
By combining equations (**1**) and (**2**) the generated piezoelectric potential is expressed as:

$$\text{Piezoelectric Potential} = K* f(T_1, T_2 \ldots\ldots T_n) \qquad (3)$$

Equation (**3**) indicates that the generated piezoelectric potential will be composed by oscillating electric field components which will correspond to the different $T_i$ of equation (**3**) which in turn correspond to the oscillating tidal components. By isolating a single electric oscillating component, equation (**3**) gives as a result of such operation the **block (3)** presented in figure (**3**).

Now, we must recall the fact that the rock fracturing, that is the generation of an earthquake, takes place at the peaks of the stress load applied on a rock formation. Consequently, it is interesting to compare the occurrence time of large EQs (**block 4**) data file to the observed piezoelectric oscillating electric field. For the implementation of this test, from the raw Earth's electric field an oscillating component is isolated by band pass (**FFT**) filtering. The selected oscillation period is **T = 1 day** which corresponds to the **K1** tidal component. Thus, **block (3)** and **block (4)** of figure (**3**) give as a schematic result the **block (5)**. The black line represents the oscillating Earth's piezoelectric component for **T = 1 day** while the red bars indicate the time of occurrence of large EQs.

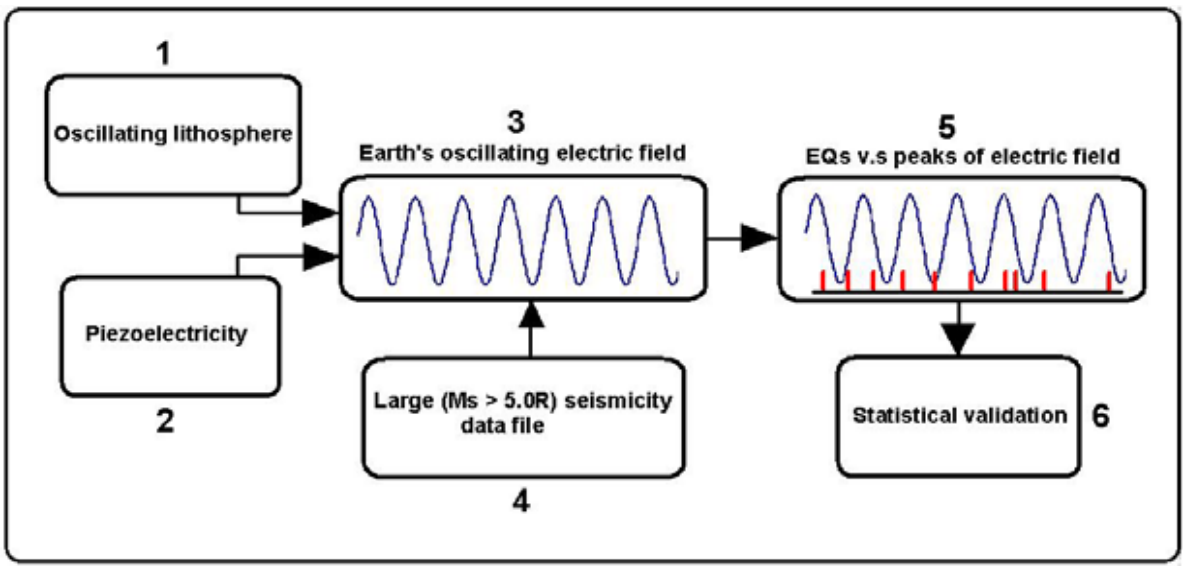

Fig. 3. Schematic presentation of the postulated physical model. The black arrows indicate the model flow direction. **1** = oscillating lithosphere due to tidal forces, **2** = activation of piezoelectricity due to lithospheric high content of quartzite, **3** = generation of Earth's oscillating electric field, **4** = Large EQs data file, **5** = correlation of large EQs (red bars) time of occurrence to amplitude peaks of the Earth's oscillating electric field, **6** = statistical validation.

The postulated model (**blocks 1 to 5**) may be either wrong or correct. Its validity will be tested statistically. If it is wrong then the occurrence time of the large EQs will not correlate to the amplitude peaks of the Earth's oscillating electric field of **T = 1 day**. On the other hand, if there is an acceptable level of correlation then the postulated model can be considered as a valid one.

3. Data analysis.

A typical oscillating field of a period of one day (**24 hours**) presents two successive amplitude peaks with a time distance between them of twelve (**12**) hours. Consequently, any physical event at any time of occurrence, within the time of one period of the oscillating field, will deviate in time from any amplitude peak of the same period for at most six (**6**) hours or 360 minutes. For the present case of a real oscillating Earth's electric field, since the band-width of the filter which was used to isolate a single oscillating component is not ideal, the derived out put will be interfered by noise of similar periods, different amplitudes and phase lags. Therefore, the six (**6**) hours maximum deviation, of any physical event from the closest amplitude peak of the oscillating field which was referred earlier, can be modified to some extent up to **450 ÷ 500** minutes. The latter is clearly shown in the **E − W** component presented in figure (**2**).

The seismic events of large magnitude (**Ms ≥ 5.0R**), which are used in the present study (for the entire study period) is **280** and were obtained from the EQ files downloaded from the National Observatory of Athens (**NOA**).

Initially, it is hypothesized that the occurrence time of large EQs is a random process compared to the amplitude peaks of the Earth's oscillating electric field (**T = 1 day**). In order to clearly demonstrate it a data set of **280** random numbers in the range of **-500** to **500** were generated by a random number generator. This random data set represents the EQs occurrence time deviations from the amplitude peaks of the oscillating electric field. Next, the random data set was statistically processed by applying occurrence frequency distribution by defining a statistical bin of **50** minutes wide. The results are shown in figure (**4**). The blue bars indicate the population of each bin, the red line represents the trend of the frequency distribution while the black line represents the "by chance" threshold of population of the entire data set.

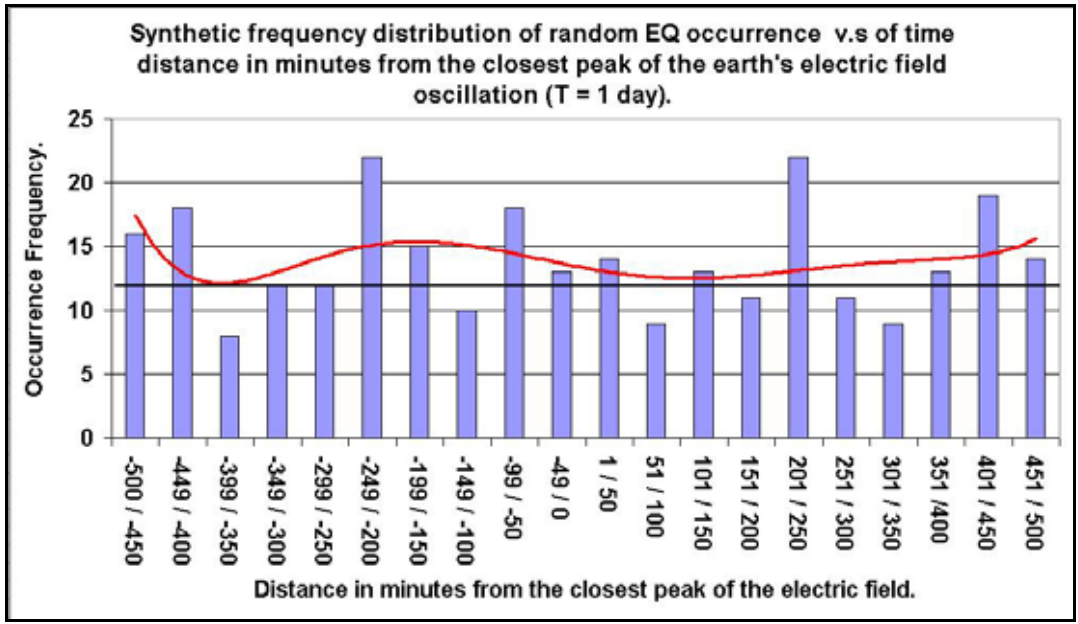

**Fig. 4.** Synthetic frequency distribution of random EQ occurrence vs. time distance in minutes from the closest amplitude peak of the earth's electric field oscillation (**T = 1 day**). The blue bars indicate the population of each bin, the red line represents the trend of the frequency distribution while the black line represents the "by chance" threshold of the population of the entire data set.

Figure (4) shows that for the entire study space (-500 ÷ +500) the bin population fluctuates randomly around the "by chance" level. Some large peaks are randomly located in the same space while the added trend shows no statistical significant importance.

The earth's electric field data which will be analyzed were registered by two monitoring sites. The first is the one located in Athens (**ATH**) and the second is located in Pyrgos (**PYR**). Both monitoring sites register the **N – S** and the **E – W** components of the Earth's electric field. Therefore, the four data sets (**ATH: EW / NS** and **PYR: EW / NS**) will be separately analyzed. The results of the analysis are shown in the following graphs:

**ATH monitoring site.**

The comparison of the EQ data file (**280 EQs**) to the **E – W** Earth's oscillating electric field component has shown that **145 EQs** have occurred closer to an **E - W** amplitude peak of the oscillating electric field while **135 EQs** closer to a **N – S** one. The **E – W** and **N – S** frequency distributions of the time distance of each EQ from the closest electric field amplitude peak are shown in the following figures (**5**) and (**6**).

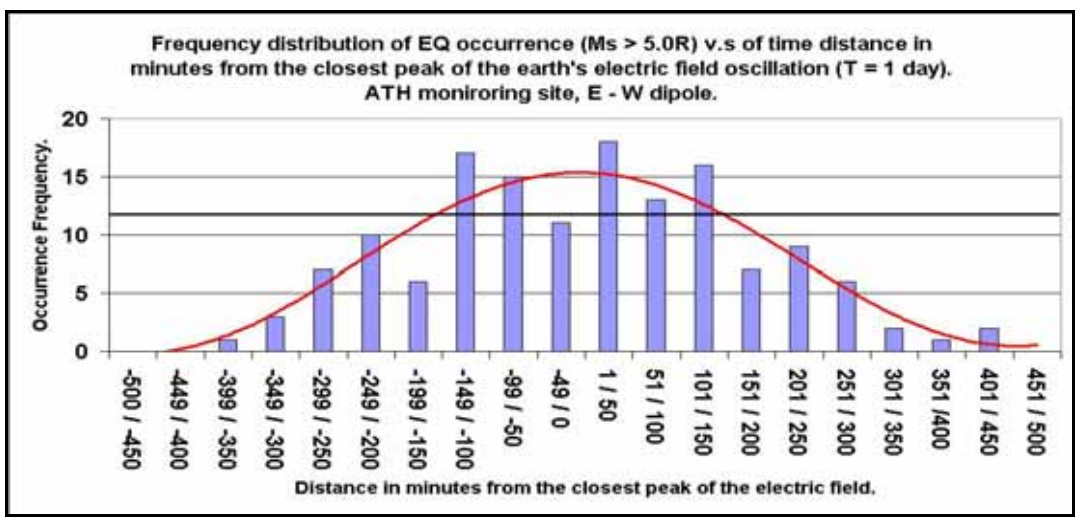

**Fig. 5.** Frequency distribution of the time distance of the large EQs (**Ms ≥ 5.0R**) from the closest amplitude peak of the Earth's oscillating electric field (**E – W** component, **ATH** monitoring site). The blue bars indicate the population of each bin, the black horizontal line represents the "by chance" threshold level while the red line indicates the corresponding trend.

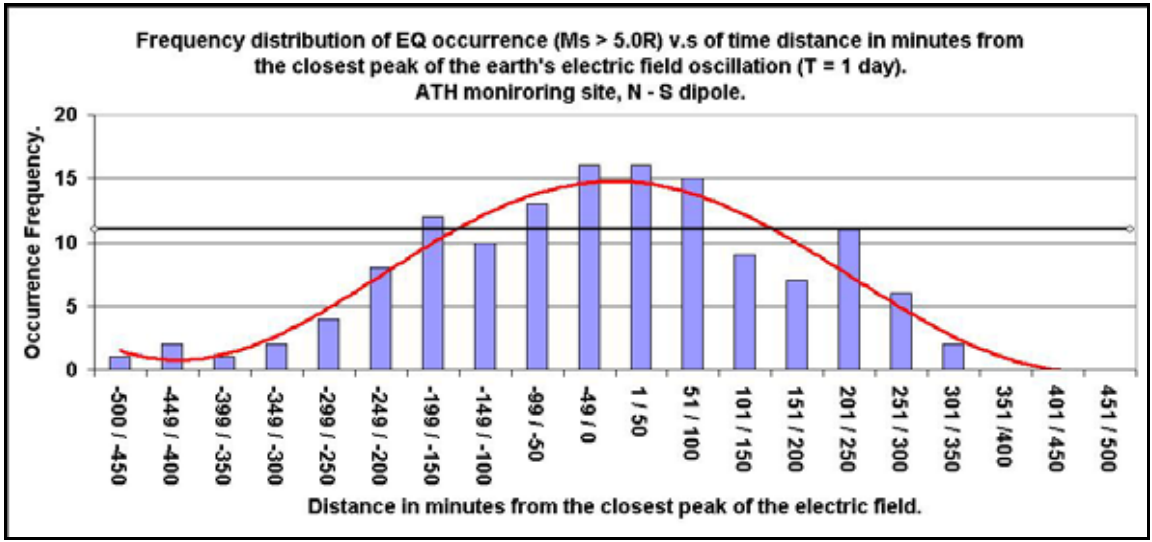

Fig. 6. Frequency distribution of the time distance of the large EQs (Ms ≥ 5.0R) from the closest amplitude peak of the Earth's oscillating electric field (N – S component, ATH monitoring site). The blue bars indicate the population of each bin, the black horizontal line represents the "by chance" threshold level while the red line indicates the corresponding trend.

In both graphs (5 and 6) a typical normal distribution is observed. The center values are well above the "by chance" level while the "tale" values are quite below it. In the following figure (7) both E – W and N – S data are combined in a single graph.

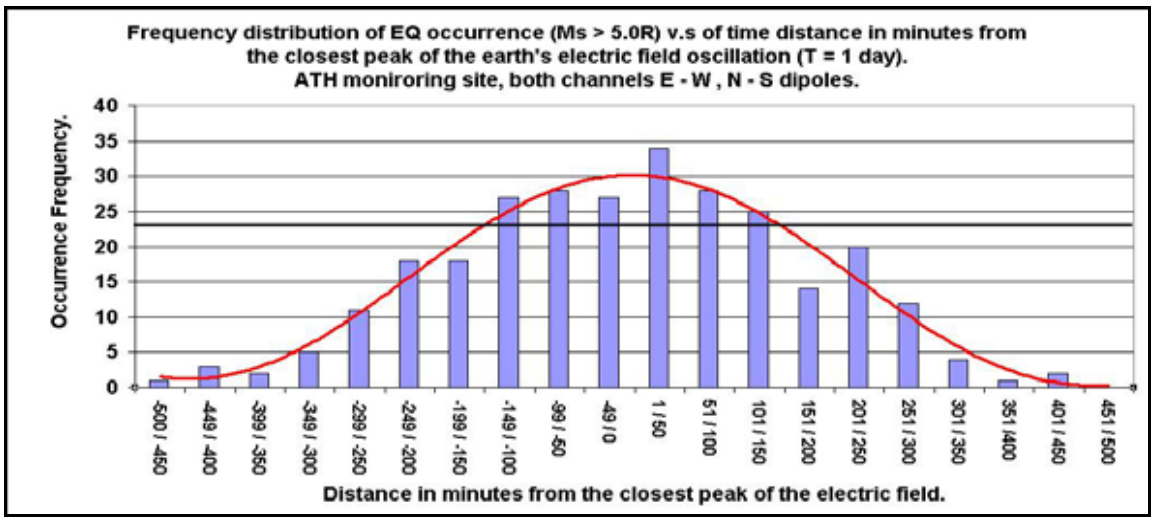

Fig. 7. Frequency distribution of the time distance of the large EQs (Ms ≥ 5.0R) from the closest amplitude peak of the Earth's oscillating electric field (E – W and N – S components, ATH monitoring site). The blue bars indicate the population of each bin, the black horizontal line represents the "by chance" threshold level while the red line indicates the corresponding trend.

**PYR monitoring site.**

The comparison of the EQ data file (**280 EQs**) to the **E – W** Earth's oscillating electric field component has shown that **149 EQs** have occurred closer to an **E - W** amplitude peak of the oscillating electric field while **131 EQs** closer to a **N – S** one. The **E – W** and **N – S** frequency distributions of the time distance of each EQ from the closest oscillating electric field amplitude peak are shown in the following figures (8) and (9).

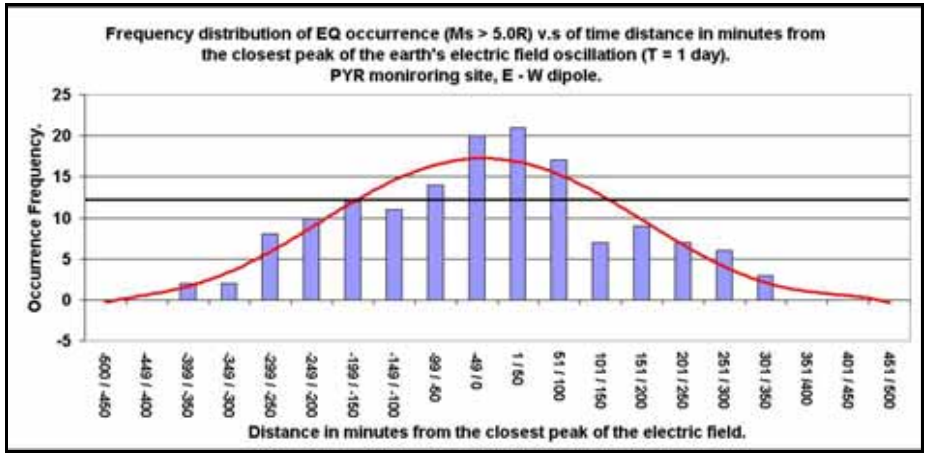

**Fig. 8.** Frequency distribution of the time distance of the large EQs (**Ms ≥ 5.0R**) from the closest amplitude peak of the Earth's oscillating electric field (**E – W** component, **PYR** monitoring site). The blue bars indicate the population of each bin, the black horizontal line represents the "by chance" threshold level while the red line indicates the corresponding trend.

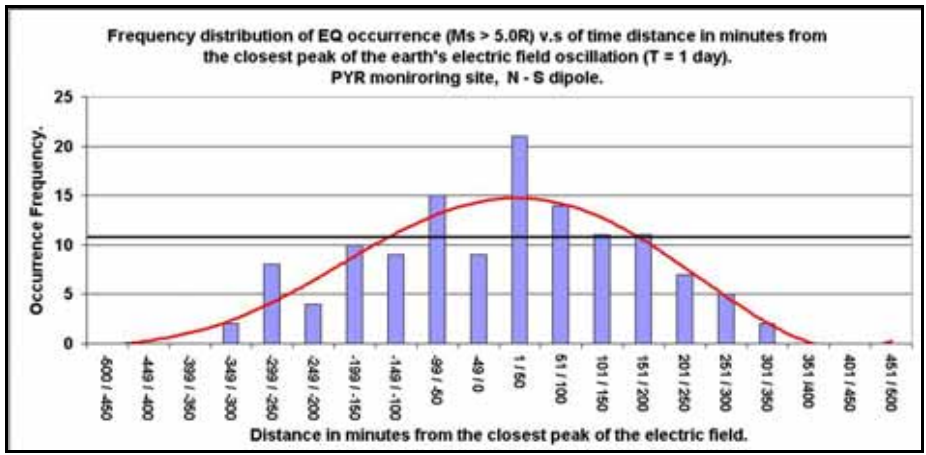

**Fig. 9.** Frequency distribution of the time distance of the large EQs (**Ms ≥ 5.0R**) from the closest amplitude peak of the Earth's oscillating electric field (**N – S** component, **PYR** monitoring site). The blue bars indicate the population of each bin, the black horizontal line represents the "by chance" threshold level while the red line indicates the corresponding trend.

In both graphs (**8** and **9**) a typical normal distribution is observed. The center values are well above the "by chance" level while the "tale" values are quite below it. In the following figure (**10**) both **E – W** and **N – S** data are combined in a single graph.

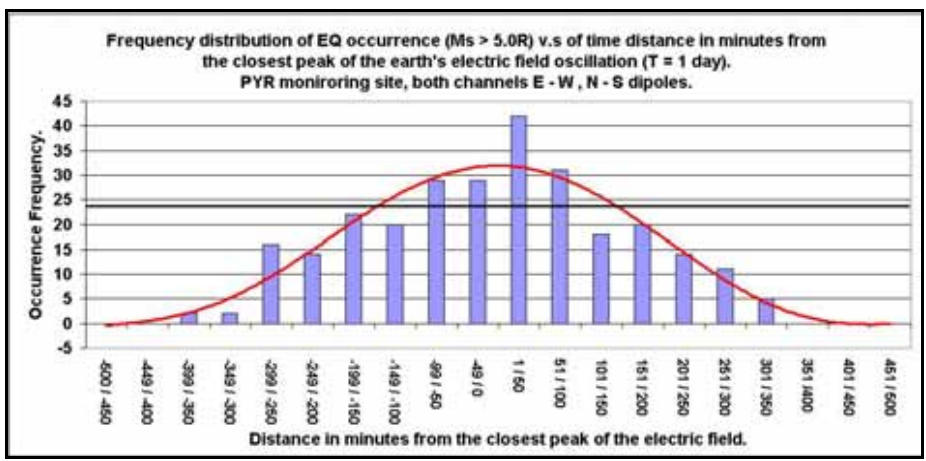

**Fig. 10.** Frequency distribution of the time distance of the large EQs (**Ms ≥ 5.0R**) from the closest amplitude peak of the Earth's oscillating electric field (**E – W** and **N – S** components, **PYR** monitoring site). The blue bars indicate the population of each bin, the black horizontal line represents the "by chance" threshold level while the red line indicates the corresponding trend.

## 4. Discussion - Conclusions.

It is hypothesized, in this work, that the tidal forces, which act upon the lithosphere, activate large scale piezoelectricity. Therefore, it is expected a close relationship between the activated piezoelectric oscillating field, the tidally generated lithospheric oscillating stress load and consequently the time of occurrence of large EQs. The expected relationship at first was assumed as a random one. A synthetic random data set was analyzed in the time interval between two successive amplitude peaks of the oscillating Earth's field component of **T = 1 day**. The random model (**fig. 4**) presents a rather constant behavior independently from the time distance of the EQ occurrence time from the amplitude peaks. The frequency population, for all used bins, fluctuates closely around the "by chance" level that was calculated for the entire data set (**280 EQs**).

Analyzing the real data (Large EQs and Earth's oscillating electric field) it is expected statistically that, 50% of the EQs will normally occur close to each peak from the two present in a day's period of time of the Earth's oscillating electric field. The actual analysis has shown that for the case of **ATH** monitoring site 145 EQs (145/280 = 51.8%) occurred closer to a peak of the **E – W** component of the Earth's oscillating electric field while **138 EQs** (138/280 = 48.2%) occurred closer to the **N – S** one. For the case of **PYR** monitoring site the corresponding values are: for the **E – W** component **149 EQs** (149/280 = 53.2%) occurred closer to the peak of the **E – W** component of the Earth's oscillating electric field while **131 EQs** (131/280 = 46.8%) occurred closer to the **N – S** one. In both cases (**ATH** and **PYR**) and for both (**N – S, E – W**) components the data set behaved quite well (very close to **50%**).

The postulated physical model of figure (**3**) was statistically tested against real data obtained not only from two different monitoring sites (**ATH, PYR**) but also from two (**N - S** and **E – W**) different components of the electric field observed at each monitoring site. Figures (**5, 6, 8, 9**) show that the deviation of the time of occurrence of large EQs from the closest amplitude peak of the Earth's oscillating electric field follows a typical normal distribution. Therefore, the postulated model is physically justified although the seismic events are distributed almost equally tor the **E – W** and **N – S** components of the Earth's oscillating electric field. Very similar results are shown in figures (**7, 10**) where the results of both directions have been merged in a single graph. In all graphs of figures (**5, 6, 7, 8, 9, 10**) the normal distribution graph peaks at the zero-deviation value. This result fosters the hypothesis that the tidal forces can very well be considered as an activating mechanism for the generation of the Earth's oscillating electric field through large scale piezoelectricity (Thanassoulas, 2008b). The latter is explained by the fact that the tidal forces are the main cause for the generation of the oscillating lithospheric stress-load which in turn controls the time of occurrence of large EQs (Thanassoulas et al. 2011).

A very interesting feature of the graphs of figures (**7, 10**) is the presence of a distinct population peak, observed at both monitoring sites, at the bin **1 ÷ 50** minutes. That feature is indicated by a red arrow in the following figure (**11**) for **ATH** and (**12**) for **PYR** monitoring sites. It indicates that the large EQs "prefer" to occur after the amplitude peak of the oscillating electric field and consequently, following the postulated physical model, after the corresponding stress-load peak. It is clear that there is a delay of the occurrence of a large EQ compared to the time of occurrence of the maximum applied lithospheric stress load. The latter, most probably, is related to the initiation time required for the occurrence of a large EQ. The same level of delay (up to **50** minutes) had been reported by Thanassoulas et al. (2010) for two past large EQs in Greece. The first one, of Skyros Island in Greece (**26/7/2001, Ms = 6.1 R**), was delayed for **41** minutes from the corresponding tidal peak while the second one, of Kythira in Greece (**08/01/2006, Ms = 6.9R**) was delayed for **43** minutes from the corresponding tidal peak.

In the following figures (**11, 12**) the specific observed peaks, for both monitoring sites, are indicated by a red arrow.

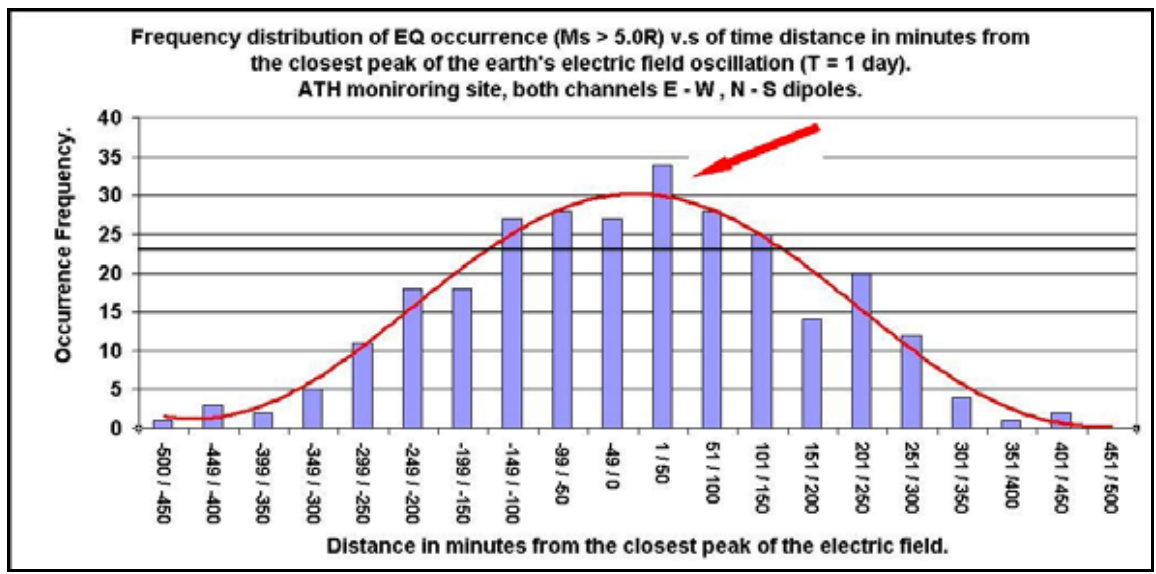

Fig. 11. Frequency distribution of the time distance of the large EQs (**Ms ≥ 5.0R**) from the closest amplitude peak of the Earth's oscillating electric field (**E – W** and **N – S** components, **ATH** monitoring site). The blue bars indicate the population of each bin, the black horizontal line represents the "by chance" threshold level while the red line indicates the corresponding trend. The red arrow indicates the "bin" **1 ÷ 50** minutes.

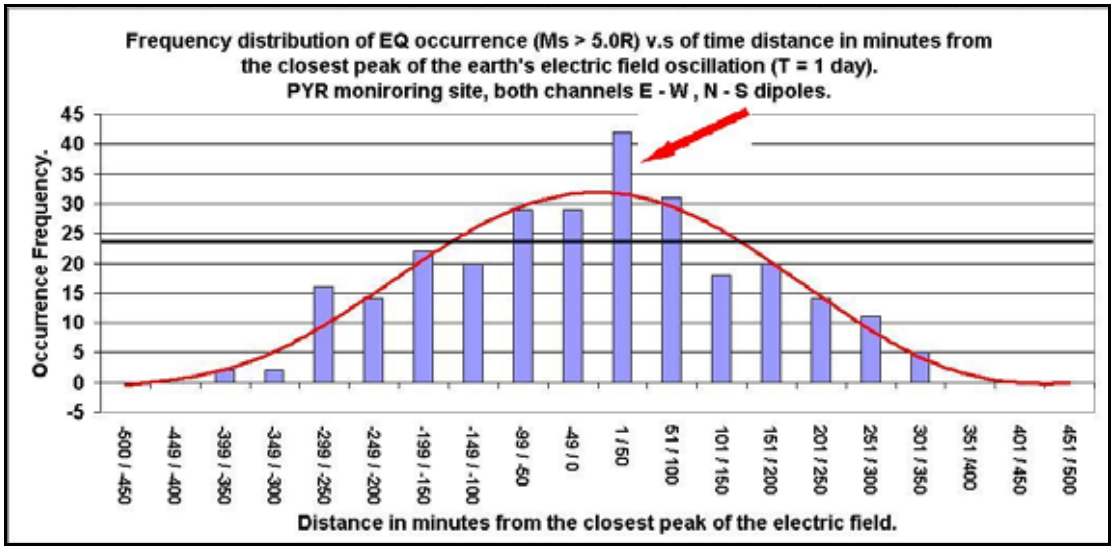

**Fig. 12.** Frequency distribution of the time distance of the large EQs (Ms ≥ 5.0R) from the closest amplitude peak of the Earth's oscillating electric field (E – W and N – S component, PYR monitoring site). The blue bars indicate the population of each bin, the black horizontal line represents the "by chance" threshold level while the red line indicates the corresponding trend. The red arrow indicates the "bin" **1 ÷ 50 minutes**.

In conclusion, the postulated physical model is justified by the results of this analysis. It seems that the large EQs do occur in close relation to the Earth's oscillating (**T = 1 day**) electric field amplitude peaks. A very similar work in results, concerning the oscillating Earth's electric field of **T = 365 days**, was presented by Thanassoulas et al. (2009b). It was shown in that work that the large EQs occurrence time coincided to the amplitude peaks of the oscillating Earth's electric field of **T = 365 days**. That electric field is generated by the deformation of the Earth's lithosphere due to its orbit around the Sun.

What is important in this physical model is its use in short-term earthquake prediction. By the study of other earthquake precursors i.e. seismic electric signals (**SES**), sudden increase of the amplitude of the Earth's oscillating electric field, presence of very long period (**VLP**) signals, strange attractor like earthquake precursors, tidal analysis at **T = 1 / 14 days**, it is possible to narrow the time window for the occurrence of a large EQ down to a day (see examples at www.earthquakeprediction.gr). The postulated model suggests only two (statistically) valid times within a day that are mostly favorable for the occurrence of a large EQ. Since two components (**E - W** and **N – S**) of the oscillating electric field are studied, then, only four discrete favorable times can be identified within a day for a single monitoring site. A recent example of a large EQ that did occur during the preparation of this work is presented in the following figures (13) and (14). The EQ occurred on August 7th, 2011 near Nafpactos town with a magnitude **Ms = 5.3R**. The specific EQ was preceded by an **SES** (train like short electric pulses) generated on August 1st, August 2011 which was detected by both **PYR** and **ATH** monitoring sites. Moreover, a strange attractor like precursor was generated by the Earth's oscillating (**T = 14 days**) electric field that lasted for two days before the EQ occurrence (see [www.earthquakeprediction.gr](www.earthquakeprediction.gr)) . Consequently, it was a matter of a couple of days for the large EQ to occur.

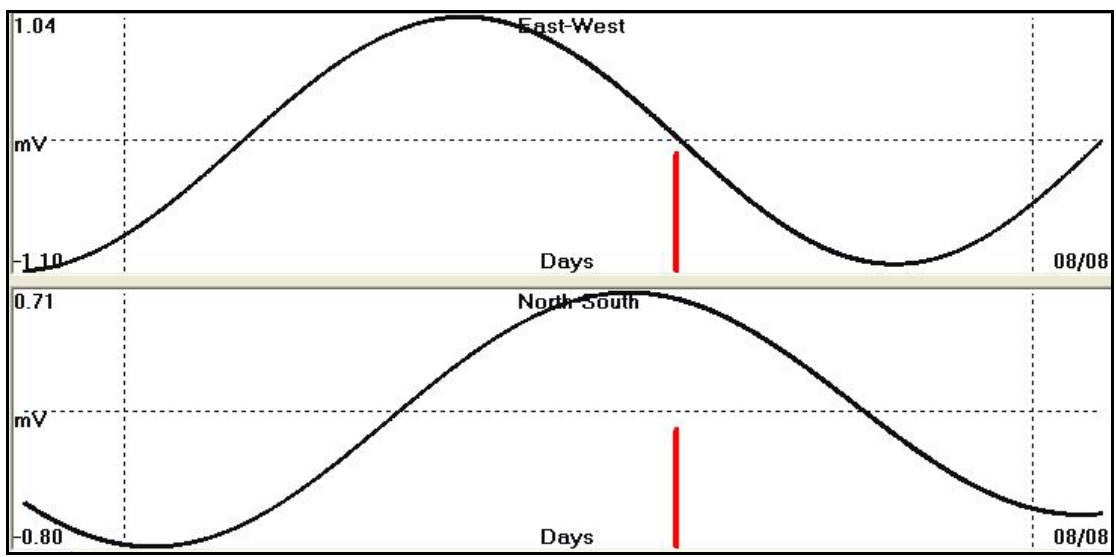

**Fig. 13.** At **ATH** monitoring site, the time deviation of the EQ occurrence (red bar) from **N – S** peak is: **dt = +75 minutes**.

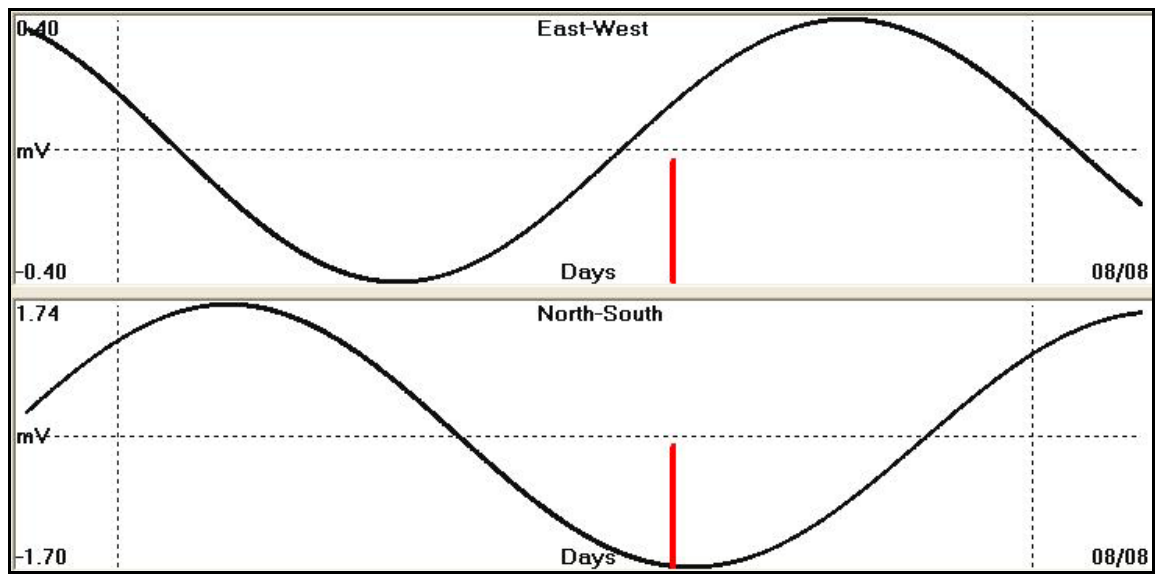

**Fig. 14. At PYR monitoring site, the time deviation of the EQ occurrence (red bar) from N – S peak is: dt = -26 minutes.**

From two (2) monitoring sites, eight (8) discrete times within a day can be defined, two for each Earth's oscillating electric component at each monitoring site. By constraining these times with the daily tidal variation it is possible to define very accurately the time of occurrence of a large EQ (Thanassoulas, 2007).

Finally, it must be pointed out that the Earth's electric field contains, among others, a lot of useful information related to the generation of large EQs. What is really needed is the appropriate methodology so that to make full use of it towards a practical earthquake prediction scheme.

## 5. References.

**www.earthquakeprediction.gr**